\documentclass[12pt]{article}
\usepackage{amsfonts}
\usepackage{amsmath}
\usepackage{amssymb}

\input{tcilatex}
\begin{document}

\begin{center}
\smallskip \ 

\textbf{GEOMETRIC STRUCTURE}

\smallskip \ 

\textbf{OF HIGHER-DIMENSIONAL SPHERES}

\smallskip \ 

G. Avila$^{a}$ \footnote{%
gavilasierra@posgrado.cifus.uson.mx} S. J. Castillo$^{a}$ \footnote{%
semiconductores@difus.uson.mx} and J. A. Nieto$^{b}$ \footnote{%
nieto@uas.edu.mx, janieto1@asu.edu}

\smallskip \ 

\smallskip \ 

\smallskip \ 

$^{a}$\textit{Departameto de Investigaci\'{o}n en F\'{\i}sica de la
Universidad}

\textit{\textit{de Sonora, C.P. 83000, }Hermosillo, Sonora, M\'{e}xico}

\smallskip \ 

$^{b}$\textit{Facultad de Ciencias F\'{\i}sico-Matem\'{a}ticas de la
Universidad Aut\'{o}noma}

\textit{de Sinaloa, C.P. 80010, Culiac\'{a}n, Sinaloa, M\'{e}xico}

\smallskip \ 

\smallskip \ 

\textbf{Abstract}
\end{center}

We explain in some detail the geometric structure of spheres in any
dimension. Our approach may be helpful for other homogeneous spaces (with
other signatures) such as the de Sitter and anti-de Sitter spaces. We apply
the procedure to the recently proposed division-algebras/Poincar\'{e}%
-conjecture correspondence. Moreover, we explore the possibility of a
connection between $N$-qubit system and the Hopf maps. We also discuss the
possible links of our work with squashed-spheres in supergravity and
pseudo-spheres in oriented matroid theory.

\bigskip \ 

\bigskip \ 

\bigskip \ 


Pacs numbers: 04.20.Gz, 04.60.-Ds, 11.30.Ly

October 30., 2013

\newpage

\noindent \textbf{1. Introduction}

\smallskip \ 

\noindent Higher-dimensional spheres $S^{d}$ are of great importance in both
mathematics and physics. In mathematics, for instance, one can find the
statement that the only parallelizable spheres over the real are $%
S^{1},S^{3} $ and $S^{7}$ (see Refs. [1]-[10]) and also that any\emph{\ }%
simply\emph{\ }connected\emph{\ }compact manifold\emph{\ }over the real\emph{%
\ }must be homeomorphic to $S^{d}$ (generalized Poincar\'{e}-conjecture)
(see Refs. [11]-[15]). Topologically, the relevance of several spheres
emerges trough the Hopf maps $S^{3}\overset{S^{1}}{\longrightarrow }S^{2}$, $%
S^{7}\overset{S^{3}}{\longrightarrow }S^{4}$ and $S^{15}\overset{S^{7}}{%
\longrightarrow }S^{8}$ [4], with fibers $S^{1},S^{3}$ and $S^{7}$
respectively. Of course these fiber spaces are deeply related to the normed
division algebras; real numbers, complex numbers, quaternions and octonions
(see Ref. [10]). Surprisingly, there is also an intriguing relation between $%
N$-qubit theory and the spheres $S^{1},S^{3}$ and $S^{7}$ [16]-[18] (see
also Ref. [19]). Moreover, the sphere $S^{8}$ is of great relevance in Bott
periodicity theorem (see Ref. [10] and references therein). In physics, one
meets with $S^{3}$ in the Friedmann-Robertson-Walker cosmological model (see
Ref. [20] and references therein), while in supergravity and superstring
compactification (see Ref. [21] and references therein) one learns that one
of the most interesting candidate for a realistic Kaluza-Klein theory is $%
S^{7}$ or the corresponding squashed sphere $\mathcal{S}^{7}$ [22].

In this article, we would like to explain the geometry structure of any
higher-dimensional sphere $S^{d}$. Our method is straightforward and can be
applied to any spacetime with $t$-time and $s$-space signature. We explain
how to obtain the de Sitter space in spherical coordinates. Moreover, as an
application we link our approach with the division-algebras and the Poincar%
\'{e}-conjecture. We also mention some generalizations of $S^{d}$ such as
squashed spheres in supergravity and pseudo spheres in oriented matroid
theory.

\bigskip \ 

\noindent \textbf{2. Spheres as a constrained system}

\smallskip \ 

\noindent Let us consider coordinate $x^{A}$ with $A=1,2,...,d+1$. We define
the $S^{d}$ sphere, with constant radius $r_{0}$, through the constraint

\begin{equation}
x^{i}x^{j}\delta _{ij}+x^{d+1}x^{d+1}=r_{0}^{2},  \tag{1}
\end{equation}%
where the $\delta _{ij}$ is the Kronecker delta and $i=1,2,...,d$.

We shall be interested in the line element

\begin{equation}
ds^{2}\equiv dx^{A}dx^{B}\delta _{AB}=dx^{i}dx^{j}\delta
_{ij}+dx^{d+1}dx^{d+1}.  \tag{2}
\end{equation}%
From (1) one obtains

\begin{equation}
x^{d+1}=\pm (r_{0}^{2}-x^{i}x^{j}\delta _{ij})^{1/2}.  \tag{3}
\end{equation}%
Thus, taking the differential of (3) gives

\begin{equation}
dx^{d+1}=-\frac{(\pm )x^{i}dx^{j}\delta _{ij}}{(r_{0}^{2}-x^{r}x^{s}\delta
_{rs})^{1/2}}.  \tag{4}
\end{equation}%
So, substituting (4) into (2) leads to

\begin{equation}
ds^{2}\equiv dx^{i}dx^{j}\delta _{ij}+\frac{(x^{i}dx^{j}\delta
_{ij})(x^{k}dx^{l}\delta _{kl})}{(r_{0}^{2}-x^{r}x^{s}\delta _{rs})}. 
\tag{5}
\end{equation}%
This expression can be rewritten as

\begin{equation}
ds^{2}\equiv dx^{i}dx^{j}g_{ij},  \tag{6}
\end{equation}%
where

\begin{equation}
g_{ij}=\delta _{ij}+\frac{x_{i}x_{j}}{(r_{0}^{2}-x^{r}x^{s}\delta _{rs})}. 
\tag{7}
\end{equation}%
Here, we used the expression $x^{k}\delta _{ki}=x_{i}$.

By using (7) we shall calculate the Christoffel symbols

\begin{equation}
\Gamma _{kl}^{i}=\frac{1}{2}g^{ij}(g_{jk,l}+g_{jl,k}-g_{kl,j}),  \tag{8}
\end{equation}%
and the Riemann tensor

\begin{equation}
R_{jkl}^{i}=\partial _{k}\Gamma _{jl}^{i}-\partial _{l}\Gamma
_{jk}^{i}+\Gamma _{mk}^{i}\Gamma _{jl}^{m}-\Gamma _{ml}^{i}\Gamma _{jk}^{m}.
\tag{9}
\end{equation}%
First, let us observe that

\begin{equation}
g_{jk,l}=\frac{1}{(r_{0}^{2}-x^{p}x^{q}\delta _{pq})}[\delta
_{jl}x_{k}+\delta _{kl}x_{j}+\frac{2x_{j}x_{k}x_{l}}{(r_{0}^{2}-x^{r}x^{s}%
\delta _{rs})}].  \tag{10}
\end{equation}%
One finds that the Christoffel symbol (8) becomes

\begin{equation}
\Gamma _{kl}^{i}=\frac{1}{(r_{0}^{2}-x^{r}x^{s}\delta _{rs})}%
g^{ij}(g_{kl}x_{j}).  \tag{11}
\end{equation}%
But since

\begin{equation}
g^{ij}=(\delta ^{ij}-\frac{x^{i}x^{j}}{r_{0}^{2}}),  \tag{12}
\end{equation}%
one finds that

\begin{equation}
\Gamma _{kl}^{i}=\frac{g_{kl}x^{i}}{r_{0}^{2}}.  \tag{13}
\end{equation}%
By substituting this result into (9) yields

\begin{equation}
\begin{array}{c}
R_{ijkl}=g_{im}R_{jkl}^{m}=g_{im}[\partial _{k}(\frac{g_{jl}x^{m}}{r_{0}^{2}}%
)-\partial _{l}(\frac{g_{jk}x^{m}}{r_{0}^{2}})+(\frac{g_{nk}x^{m}}{r_{0}^{2}}%
)(\frac{g_{jl}x^{n}}{r_{0}^{2}}) \\ 
\\ 
-(\frac{g_{nl}x^{m}}{r_{0}^{2}})(\frac{g_{jk}x^{n}}{r_{0}^{2}})].%
\end{array}
\tag{14}
\end{equation}%
Simplifying (14) gives%
\begin{equation}
R_{ijkl}=g_{im}[\partial _{k}(\frac{g_{jl}x^{m}}{r_{0}^{2}})-\partial _{l}(%
\frac{g_{jk}x^{m}}{r_{0}^{2}})+\frac{1}{r_{0}^{4}}%
x^{m}x^{n}(g_{nk}g_{jl}-g_{nl}g_{jk})].  \tag{15}
\end{equation}%
Now, taking the derivatives in the first and second terms in (15) one sees
that%
\begin{equation}
R_{ijkl}=\frac{1}{r_{0}^{2}}g_{im}[(g_{jl}\delta _{k}^{m}-g_{jk}\delta
_{l}^{m})+x^{m}(g_{jl,k}-g_{jk,l})+\frac{1}{r_{0}^{2}}%
x^{m}x^{n}(g_{nk}g_{jl}-g_{nl}g_{jk})].  \tag{16}
\end{equation}%
Using (10) one notes that (16) becomes

\begin{equation}
\begin{array}{c}
R_{ijkl}=\frac{1}{r_{0}^{2}}(g_{ik}g_{jl}-g_{il}g_{jk})+\frac{g_{im}x^{m}}{%
r_{0}^{2}(r_{0}^{2}-x^{r}x^{s}\delta _{rs})}(\delta _{jk}x_{l}-\delta
_{jl}x_{k}) \\ 
\\ 
+\frac{g_{im}}{r_{0}^{4}}x^{m}x^{n}(g_{nk}g_{jl}-g_{nl}g_{jk}).%
\end{array}
\tag{17}
\end{equation}%
But, one finds that

\begin{equation}
\begin{array}{c}
\frac{1}{r_{0}^{2}(r_{0}^{2}-x^{r}x^{s}\delta _{rs})}(\delta
_{jk}x_{l}-\delta _{jl}x_{k})+\frac{1}{r_{0}^{4}}%
x^{n}(g_{nk}g_{jl}-g_{nl}g_{jk}) \\ 
\\ 
=\frac{1}{r_{0}^{2}(r_{0}^{2}-x^{r}x^{s}\delta _{rs})}(\delta
_{jk}x_{l}-\delta _{jl}x_{k})+\frac{1}{r_{0}^{4}}[x_{k}(1+\frac{1}{%
(r_{0}^{2}-x^{r}x^{s}\delta _{rs})}x_{n}x^{n})\delta _{jl} \\ 
\\ 
-x_{l}(1+\frac{1}{(r_{0}^{2}-x^{r}x^{s}\delta _{rs})}x_{n}x^{n})\delta _{jk}]
\\ 
\\ 
=\frac{1}{r_{0}^{2}(r_{0}^{2}-x^{r}x^{s}\delta _{rs})}(\delta
_{jk}x_{l}-\delta _{jl}x_{k})+\frac{1}{r_{0}^{2}(r_{0}^{2}-x^{r}x^{s}\delta
_{rs})}(\delta _{jl}x_{k}-\delta _{jk}x_{l})=0.%
\end{array}
\tag{18}
\end{equation}%
Thus, (17) is reduced to

\begin{equation}
R_{ijkl}=\frac{1}{r_{0}^{2}}(g_{ik}g_{jl}-g_{il}g_{jk}).  \tag{19}
\end{equation}%
We recognize in this expression the typical form of the Riemann tensor for
any homogeneous space.

From (19) one learns that the Ricci tensor $R_{jl}=g^{ik}R_{ijkl}$ is given
by

\begin{equation}
R_{jl}=\frac{(d-1)g_{jl}}{r_{0}^{2}},  \tag{20}
\end{equation}%
while the scalar curvature $R=g^{jl}R_{jl}$ becomes

\begin{equation}
R=\frac{d(d-1)}{r_{0}^{2}}.  \tag{21}
\end{equation}%
Usually the theory is normalized in the sense of setting $r_{0}^{2}=\frac{1}{%
k}=1$. In this case (21) leads to

\begin{equation}
R=kd(d-1).  \tag{22}
\end{equation}%
As we shall explain in the next section the above procedure can be
generalized to no compact spacetime. In such case one obtains that (22) can
be generalized in such a way that $k=\{-1,0,1\}$.

\bigskip \ 

\noindent \textbf{3. Higher dimensional De Sitter space-time}

\smallskip \ 

Now suppose that instead of the constraint (1) we have%
\begin{equation}
x^{i}x^{j}\eta _{ij}+x^{d+1}x^{d+1}=r_{0}^{2},  \tag{23}
\end{equation}%
where we changed the Euclidean metric $\delta _{ij}=(1,1,...,1)$ by the
Minkowski metric $\eta _{ij}=diag(-1,1,...,1)$. Note that in this case the
constant $r_{0}^{2}$ can be positive, negative or zero. Accordingly, the
line element (2) is now given by

\begin{equation}
ds^{2}\equiv dx^{A}dx^{B}\delta _{AB}=dx^{i}dx^{j}\eta
_{ij}+dx^{d+1}dx^{d+1}.  \tag{24}
\end{equation}%
It is not difficult to see that all steps to calculate the Christoffel
symbols and the Riemann tensor components of the previous section are
exactly the same. At the end, it can be shown that such quantities are

\begin{equation}
\Gamma _{kl}^{i}=\frac{g_{kl}x^{i}}{r_{0}^{2}}  \tag{25}
\end{equation}%
and

\begin{equation}
R_{ijkl}=\frac{1}{r_{0}^{2}}(g_{ik}g_{jl}-g_{il}g_{jk}),  \tag{26}
\end{equation}%
respectively. Here, the metric $g_{ij}$ is now given by

\begin{equation}
g_{ij}=\eta _{ij}+\frac{x_{i}x_{j}}{(r_{0}^{2}-x^{r}x^{s}\eta _{rs})}. 
\tag{27}
\end{equation}%
where $x_{i}\equiv \eta _{ij}x^{j}$.

It is worth mentioning that one can even consider a flat metric $\eta
_{ij}=diag(-1,...,-1,....1,1)$ with $t$-times and $s$-space coordinates and
the procedure is exactly the same, that is, equations (23)-(27) are exactly
the same, with the exception that now one must take the corresponding flat
metric $\eta _{ij}$.

Just to show that our result agree with the de Sitter space-time in
spherical coordinates let us consider the reduced spacetime

\begin{equation}
ds^{2}\equiv dx^{i}dx^{j}g_{ij},  \tag{28}
\end{equation}%
obtained by using (23). Substituting (27) into (28) yields

\begin{equation}
ds^{2}\equiv (\eta _{ij}+\frac{x_{i}x_{j}}{(r_{0}^{2}-x^{r}x^{s}\eta _{rs})}%
)dx^{i}dx^{j}.  \tag{29}
\end{equation}%
This expression can be rewritten as

\begin{equation}
ds^{2}\equiv \frac{1}{(r_{0}^{2}-x^{r}x^{s}\eta _{rs})}%
[(r_{0}^{2}-x^{m}x^{n}\eta _{mn})\eta _{ij}+x_{i}x_{j}]dx^{i}dx^{j}. 
\tag{30}
\end{equation}

By expanding $x^{m}x^{n}\eta _{mn}=-x^{0}x^{0}+x^{a}x^{b}\delta _{ab}$, with 
$a,b$ running from $1$ to $d-1$, one learns that (30) leads to

\begin{equation}
\begin{array}{c}
ds^{2}\equiv \frac{1}{(r_{0}^{2}+x^{0}x^{0}-x^{e}x^{f}\delta _{ef})}%
[(r_{0}^{2}+x^{0}x^{0}-x^{a}x^{b}\delta
_{ab})(-dx^{0}dx^{0}+dx^{c}dx^{d}\delta _{cd}) \\ 
\\ 
+x^{0}x^{0}dx^{0}dx^{0}-2x^{0}dx^{0}x^{a}dx^{b}\delta
_{ab}+x^{a}x^{c}dx^{b}dx^{d}\delta _{ab}\delta _{cd}].%
\end{array}
\tag{31}
\end{equation}%
Now, considering that $r^{2}=x^{a}x^{b}\delta _{ab}$ one finds that (31) can
be written as

\begin{equation}
\begin{array}{c}
ds^{2}\equiv \frac{1}{(r_{0}^{2}+x^{0}x^{0}-r^{2})}%
[(r_{0}^{2}+x^{0}x^{0}-r^{2})(-dx^{0}dx^{0}+dr^{2}+r^{2}d\Omega ^{d-2}) \\ 
\\ 
+x^{0}x^{0}dx^{0}dx^{0}-2x^{0}dx^{0}rdr+r^{2}dr^{2}].%
\end{array}
\tag{32}
\end{equation}%
where, $d\Omega ^{d-2}$ is a volume element in $d-2$ dimensions. This can be
simplified in the form

\begin{equation}
\begin{array}{c}
ds^{2}\equiv \frac{1}{(r_{0}^{2}+x^{0}x^{0}-r^{2})}%
[-(r_{0}^{2}-r^{2})dx^{0}dx^{0}+(r_{0}^{2}+x^{0}x^{0})dr^{2}-2x^{0}dx^{0}rdr]
\\ 
\\ 
+r^{2}d\Omega ^{d-2}.%
\end{array}
\tag{33}
\end{equation}

Now, consider the change of variable%
\begin{equation}
x^{0}=f(t)(r_{0}^{2}-r^{2})^{1/2}.  \tag{34}
\end{equation}%
One has

\begin{equation}
dx^{0}=f^{\prime }(t)(r_{0}^{2}-r^{2})^{1/2}dt-\frac{f(t)rdr}{%
(r_{0}^{2}-r^{2})^{1/2}}.  \tag{35}
\end{equation}%
Consequently, one obtains

\begin{equation}
dx^{0}dx^{0}=f^{\prime 2}(t)(r_{0}^{2}-r^{2})dt^{2}-2f^{\prime }(t)f(t)rdrdt+%
\frac{f^{2}(t)r^{2}dr^{2}}{(r_{0}^{2}-r^{2})}  \tag{36}
\end{equation}%
and%
\begin{equation}
x^{0}dx^{0}=f(t)f^{\prime }(t)(r_{0}^{2}-r^{2})dt-f^{2}(t)rdr.  \tag{37}
\end{equation}%
Substituting (34), (36) and (37) into (33) yields%
\begin{equation}
\begin{array}{c}
ds^{2}\equiv \frac{1}{(r_{0}^{2}-r^{2})(1+f^{2})}[-(r_{0}^{2}-r^{2})(f^{%
\prime 2}(t)(r_{0}^{2}-r^{2})dt^{2}-2f^{\prime }(t)f(t)rdrdt \\ 
\\ 
+\frac{f^{2}(t)r^{2}dr^{2}}{(r_{0}^{2}-r^{2})}%
)+(r_{0}^{2}+f^{2}(t)(r_{0}^{2}-r^{2}))dr^{2} \\ 
\\ 
-2(f(t)f^{\prime }(t)(r_{0}^{2}-r^{2})dt-f^{2}(t)rdr)rdr)]+r^{2}d\Omega
^{d-2}.%
\end{array}
\tag{38}
\end{equation}%
It is not difficult to see that this expression can be simplified in the form%
\begin{equation}
ds^{2}\equiv -\frac{f^{\prime 2}(t)}{(1+f^{2})}(r_{0}^{2}-r^{2})dt^{2}+\frac{%
r_{0}^{2}}{(r_{0}^{2}-r^{2})}dr^{2}+r^{2}d\Omega ^{d-2}.  \tag{39}
\end{equation}%
Now, writing $f(t)$ as

\begin{equation}
f(t)=\sinh (t/r_{0}),  \tag{40}
\end{equation}%
one finally discovers that (39) can be written as%
\begin{equation}
ds^{2}\equiv -(1-\frac{r^{2}}{r_{0}^{2}})dt^{2}+\frac{dr^{2}}{(1-\frac{r^{2}%
}{r_{0}^{2}})}+r^{2}d\Omega ^{d-2}.  \tag{41}
\end{equation}%
This expression is, of course, very useful when one considers black-holes or
cosmological models in the de Sitter (or anti-de Sitter) space.

\bigskip \ 

\bigskip \ 

\noindent \textbf{4. Division-algebras and Poincar\'{e}-conjecture
correspondence.}

\smallskip \ 

Here, we shall briefly discuss the recently proposed correspondence between
the division algebras and the Poincar\'{e} conjecture (see Ref. [15]). First
let us recall that if there exist a real division algebra then the $d$%
-dimensional sphere $S^{d}$ in $R^{d+1}$ is parallelizable [1]-[3]. It is
also known that the only parallelizable spheres are $S^{1},S^{3}$ and $S^{7}$
[4] (see also Ref. [5]). So, one can conclude that division algebras only
exist in $1,2,3$ or $8$ dimensions (see Refs. [6]-[10] and references
therein). It turns out that these theorems are deeply related to the Hopf
maps, $S^{3}\overset{S^{1}}{\longrightarrow }S^{2}$, $S^{7}\overset{S^{3}}{%
\longrightarrow }S^{4}$ and $S^{15}\overset{S^{7}}{\longrightarrow }S^{8}$
[4]. As it is mentioned in Ref. [15], focusing on $S^{3}$, it is intriguing
that none of these remarkable results have been considered in the proof of
the original Poincar\'{e} conjecture [11]-[13], which establishes that any
simply connected closed $3$-manifold $\mathcal{M}^{3}$ is homeomorphic to $%
S^{3}$. In fact, until now any proof of the Poincar\'{e} conjecture
associated with $S^{3}$ is based on the Ricci flow equation [14] (see also
Refs. [11]-[13]). But the parallelizabilty of $S^{3}$ (or any $M^{3}$
manifold) is not even mentioned. The main goal of this section is to briefly
review the link between the concept of parallelizability and the Ricci flow
equation discussed in Ref. [15]. We also explain a number of physical
scenarios where such a link may be important, including special relativity,
cosmology and Hopf maps via $N$-qubit systems (see Ref. [19] and references
therein).

Before we address the problem at hand it is worth emphasizing a number of
comments. First, it is a fact that division algebras are linked to different
physical scenarios, including, superstrings [21] and supersymmetry
[23]-[24]. Even more surprising is the fact that division algebras are also
linked to quantum information theory via the $N$-qubit theory (see Refs.
[16]-[18]). Mathematically, division algebras are also connected with
important arenas such as K-theory [6]. If a division algebra is normed then
one may also introduce the four algebras; real numbers, complex numbers,
quaternions and octonions (see Ref. [10]). On the other hand the Poincar\'{e}
conjecture seems to be useful in the discussion of various cosmological
models (see Refs. [25]-[28]) and the study of gravitational instanton theory
[29].

Let us start introducing in addition to the metric $g_{ij}(x^{c})$ given in
(7) and the Christoffel symbols (8) the totally antisymmetric torsion tensor 
$T_{kl}^{i}=-T_{kl}^{i}$. Geometric parallelizability of $S^{d}$ means the
\textquotedblleft flattening\textquotedblright \ the space in the sense that

\begin{equation}
\mathcal{R}_{jkl}^{i}(\Omega _{rs}^{m})=0,  \tag{42}
\end{equation}%
where

\begin{equation}
\mathcal{R}_{jkl}^{i}=\partial _{k}\Omega _{jl}^{i}-\partial _{l}\Omega
_{jk}^{i}+\Omega _{mk}^{i}\Omega _{jl}^{m}-\Omega _{ml}^{i}\Omega _{jk}^{m},
\tag{43}
\end{equation}%
is the Riemann curvature tensor, with

\begin{equation}
\Omega _{kl}^{i}=\Gamma _{kl}^{i}+T_{kl}^{i}.  \tag{44}
\end{equation}%
By substituting (44) into (43) one finds

\begin{equation}
R_{jkl}^{i}+D_{k}T_{jl}^{i}-D_{l}T_{jk}^{i}+T_{mk}^{i}T_{jl}^{m}-T_{ml}^{i}T_{jk}^{m}=0.
\tag{45}
\end{equation}%
Here, $D_{i}$ denotes a covariant derivative with $\Gamma _{jk}^{i}$ as a
connection and $R_{jkl}^{i}$ is given by (9).

Using the cyclic identities for $R_{jkl}^{i}$ one can obtain the formula

\begin{equation}
R_{ijkl}=T_{mij}T_{kl}^{m}-T_{m[ij}T_{k]l}^{m}.  \tag{46}
\end{equation}%
But as we showed in section 2, for a $d$-dimensional sphere $S^{d}$ with
radius $r_{0}$ we have that $R_{ijkl}$ is given by (19). Therefore, one
learns that (46) leads to

\begin{equation}
\frac{1}{r_{0}^{2}}%
(g_{ik}g_{jl}-g_{il}g_{jk})=T_{mij}T_{kl}^{m}-T_{m[ij}T_{k]l}^{m}.  \tag{47}
\end{equation}%
Contracting in (47) with $g^{ij}$ and $T_{f}^{ik}$ it leads to first and the
second Cartan-Shouten equations%
\begin{equation}
T_{i}^{kl}T_{jkl}=(d-1)r_{0}^{-2}g_{ij},  \tag{48}
\end{equation}%
and

\begin{equation}
T_{ei}^{l}T_{lj}^{f}T_{fk}^{e}=(d-4)r_{0}^{-2}T_{ijk},  \tag{49}
\end{equation}%
respectively. Durander, Gursey and Tze [30] noted that (48) and (49) are
covariant forms associated with the algebraic identities derived in normed
division algebras. It turns out that (48) and (49) can be used eventually to
prove that the only parallelizable spheres are $S^{1},S^{3}$ and $S^{7}$
[5]. In general, however, for other $d$-manifolds $M^{d}$ the expressions
(47)-(49) does not hold.

If the only condition is that $M^{d}$ is parallelizable one may start with
(46) instead of (47). Then contracting (46) with $g^{ik}$ leads to

\begin{equation}
R_{ij}=T_{i}^{kl}T_{jkl}.  \tag{50}
\end{equation}%
Here, we recall that $R_{jl}=g^{ik}R_{ijkl}$ is the Ricci tensor.

Now, we would like to generalize the key constraint (1) in the form

\begin{equation}
x^{d+1}=\varphi (x^{i}),  \tag{51}
\end{equation}%
where $\varphi $ is an arbitrary function of the coordinates $x^{i}$. In
this case, the corresponding metric $\gamma _{ij}$ associated with $M^{d}$
becomes

\begin{equation}
\gamma _{ij}=\delta _{ij}+\partial _{i}\varphi \partial _{j}\varphi , 
\tag{52}
\end{equation}%
while the inverse $\gamma ^{ij}$ is given by

\begin{equation}
\gamma ^{ij}=\delta ^{ij}-\frac{\partial ^{i}\varphi \partial ^{j}\varphi }{%
1+\partial ^{k}\varphi \partial _{k}\varphi }.  \tag{53}
\end{equation}

The Christoffel symbols become%
\begin{equation}
\Gamma _{kl}^{i}=\frac{\partial ^{i}\varphi \partial _{kl}\varphi }{%
1+\partial ^{m}\varphi \partial _{m}\varphi }.  \tag{54}
\end{equation}%
After lengthy, but straightforward computation, one discovers that the
Riemann tensor $R_{ijkl}$ obtained form (9) and (54) is%
\begin{equation}
R_{ijkl}=\frac{1}{1+\partial ^{m}\varphi \partial _{m}\varphi }(\partial
_{ik}\varphi \partial _{jl}\varphi -\partial _{il}\varphi \partial
_{jk}\varphi ).  \tag{55}
\end{equation}%
One can check that in the particular case of

\begin{equation}
\varphi =(r_{0}^{2}-x^{i}x^{j}\delta _{ij})^{1/2},  \tag{56}
\end{equation}%
the equation (19) follows from (55).

Let us now consider the Ricci flow evolution equation [14] (see also Refs.
[11]-[13] and references therein)

\begin{equation}
\frac{\partial \gamma _{ij}}{\partial t}=-2R_{ij}.  \tag{57}
\end{equation}%
In this case the metric $\gamma _{ij}(t)$ is understood as a family of
Riemann metrics on $M^{3}$. It has been emphasized that the Ricci flow
equation is the analogue of the heat equation for metrics $\gamma _{ij}$.
The central idea is that a metric $\gamma _{ij}$ associated with a simply
connected closed manifold $\mathcal{M}^{3}$ evolves according to (57)
towards a metric $g_{ij}$ for $S^{3}$. Symbolically, this means that in
virtue of (57) we have the metric evolution $\gamma _{ij}\longrightarrow
g_{ij}$, which in turn must imply the homeomorphism $\mathcal{M}%
^{3}\longrightarrow S^{3}$.

The question arises whether one can introduce the parallelizability concept
into (57). Let us assume that $\mathcal{M}^{3}$ is a parallelizable
manifold. We shall also assume that $\mathcal{M}^{3}$ is determined by the
general constraint (51). First, we observe that using (50), the Ricci
equation (57) can be written as

\begin{equation}
\frac{\partial \gamma _{ij}}{\partial t}=-2T_{i}^{kl}T_{jkl}.  \tag{58}
\end{equation}%
This is an interesting result because it means that the evolution of $\gamma
_{ij}$ may also be determined by the torsion tensor $T_{kl}^{i}$. In the
case of $S^{3}$ manifold, using (19) or (20) one obtains an Einstein type
metric

\begin{equation}
R_{ij}=\frac{2}{r_{0}^{2}}g_{ij},  \tag{59}
\end{equation}%
which implies the evolution equation

\begin{equation}
\frac{\partial g_{ij}}{\partial t}=-\frac{4}{r_{0}^{2}}g_{ij}.  \tag{60}
\end{equation}%
These type of equations are discussed extensively in references [11] and
[13]. A relevant feature from the solution is that at a large time evolution
the behavior of $g_{ij}$ is $g_{ij}(t)=(1-\frac{2}{r_{0}^{2}}t)g_{ij}(0)$,
where $g_{ij}(0)$ corresponds to an initial condition for the metric. In
this case one has $R_{ij}(t)=R_{ij}(0)$ and since $\frac{2}{r_{0}^{2}}>0$
one has uniform contraction with singularity at $t=\frac{r_{0}^{2}}{2}$ (see
Ref. [13] for details).

\bigskip \ 

\noindent \textbf{5. Stereographic projection for the }$d$\textbf{-sphere}

\smallskip \ 

Consider the stereographic projection

\begin{equation}
X^{i}=\frac{x^{i}}{1-x^{d+1}}.  \tag{61}
\end{equation}%
Here, the coordinates $x^{i}$ and $x^{d+1}$ satisfy the constraint (1), with 
$r_{0}=1$. This means that such coordinates determine a $d$-sphere with
radius $r_{0}=1$. From (1) and (61) one may derive

\begin{equation}
x^{i}=\Phi X^{i},  \tag{62}
\end{equation}%
and

\begin{equation}
x^{d+1}=\frac{1}{2}\Phi (\rho ^{2}-1).  \tag{63}
\end{equation}%
Here, we have used the definitions%
\begin{equation}
\Phi =\frac{2}{\rho ^{2}+1},  \tag{64}
\end{equation}%
and

\begin{equation}
\rho ^{2}=X^{i}X^{j}\delta _{ij}.  \tag{65}
\end{equation}

From (62) and (63) one has

\begin{equation}
dx^{i}=\Phi dX^{i}+d\Phi X^{i},  \tag{66}
\end{equation}%
and%
\begin{equation}
dx^{d+1}=\frac{1}{2}d\Phi (\rho ^{2}-1)+\Phi \rho d\rho ,  \tag{67}
\end{equation}%
respectively. One can substitute (66) and (67) either into (2) or (6) (with $%
g_{ij}$ given by (7)) and the result is the same, namely

\begin{equation}
ds^{2}=\Phi ^{2}dX^{i}dX^{j}\delta _{ij}.  \tag{68}
\end{equation}

We shall prove these results in both cases. First, substituting (66) and
(67) into (2) gives

\begin{equation}
\begin{array}{c}
ds^{2}=(\Phi dX^{i}+d\Phi X^{i})(\Phi dX^{j}+d\Phi X^{j})\delta _{ij} \\ 
\\ 
+(\frac{1}{2}d\Phi (\rho ^{2}-1)+\Phi \rho d\rho )(\frac{1}{2}d\Phi (\rho
^{2}-1)+\Phi \rho d\rho ).%
\end{array}
\tag{69}
\end{equation}%
This can be written as%
\begin{equation}
ds^{2}=\Phi ^{2}dX^{i}dX^{j}\delta _{ij}+A,  \tag{70}
\end{equation}%
where%
\begin{equation}
\begin{array}{c}
A\equiv 2\Phi \rho d\rho d\Phi +\rho ^{2}d\Phi ^{2}+\frac{1}{4}d\Phi
^{2}(\rho ^{2}-1)^{2} \\ 
\\ 
+\Phi \rho d\rho d\Phi (\rho ^{2}-1)+\Phi ^{2}\rho ^{2}d\rho ^{2}.%
\end{array}
\tag{71}
\end{equation}%
Our goal is to show that $A=0$. Considering (64) one sees that%
\begin{equation}
d\Phi =-\frac{4\rho d\rho }{(\rho ^{2}+1)^{2}},  \tag{72}
\end{equation}%
and

\begin{equation}
\Phi d\Phi =-\frac{8\rho d\rho }{(\rho ^{2}+1)^{3}}.  \tag{73}
\end{equation}%
Thus, substituting (72) and (73) into (71) one arrives to%
\begin{equation}
\begin{array}{c}
A=-\frac{16\rho ^{2}d\rho ^{2}}{(\rho ^{2}+1)^{3}}+\frac{16\rho ^{4}d\rho
^{2}}{(\rho ^{2}+1)^{4}}+\frac{4\rho ^{2}d\rho ^{2}(\rho ^{2}-1)^{2}}{(\rho
^{2}+1)^{4}} \\ 
\\ 
-\frac{8\rho ^{2}d\rho ^{2}(\rho ^{2}-1)}{(\rho ^{2}+1)^{3}}+\frac{4\rho
^{2}d\rho ^{2}}{(\rho ^{2}+1)^{2}}.%
\end{array}
\tag{74}
\end{equation}%
This expression leads to%
\begin{equation}
\begin{array}{c}
A=\frac{\rho ^{2}d\rho ^{2}}{(\rho ^{2}+1)^{4}}[-16(\rho ^{2}+1)+16\rho
^{2}+4(\rho ^{2}-1)^{2} \\ 
\\ 
-8(\rho ^{2}+1)(\rho ^{2}-1)+4(\rho ^{2}+1)^{2}].%
\end{array}
\tag{75}
\end{equation}%
At this level, it is not difficult to show that in fact one has the result $%
A=0$.

An alternative method to prove that (68) holds, it is to use (6), with (7)
as a metric. Observe first that%
\begin{equation}
ds^{2}=(\delta _{ij}+\frac{x_{i}x_{j}}{1-x^{k}x_{x}})dx^{i}dx^{j},  \tag{76}
\end{equation}%
implies

\begin{equation}
ds^{2}=(\delta _{ij}+\frac{\Phi ^{2}X_{i}X_{j}}{1-\Phi ^{2}\rho ^{2}})(\Phi
dX^{i}+d\Phi X^{i})(\Phi dX^{j}+d\Phi X^{j}),  \tag{77}
\end{equation}%
where we have used (62), (65) and (66). The expression (77) leads to

\begin{equation}
ds^{2}=\Phi ^{2}dX^{i}dX^{j}\delta _{ij}+B.  \tag{78}
\end{equation}%
Here, we used the definition

\begin{equation}
\begin{array}{c}
B\equiv 2\Phi \rho d\rho d\Phi +\rho ^{2}d\Phi ^{2} \\ 
\\ 
+\frac{1}{1-\Phi ^{2}\rho ^{2}}(\Phi ^{4}\rho ^{2}d\rho ^{2}+2\Phi ^{3}\rho
^{3}d\rho d\Phi +\Phi ^{2}\rho ^{4}d\Phi ^{2}).%
\end{array}
\tag{79}
\end{equation}%
Our aim is to show that in this case $B=0$. Let us start writing (79) as

\begin{equation}
\begin{array}{c}
B=\frac{1}{1-\Phi ^{2}\rho ^{2}}[2\Phi \rho d\rho d\Phi (1-\Phi ^{2}\rho
^{2})+\rho ^{2}d\Phi ^{2}(1-\Phi ^{2}\rho ^{2}) \\ 
\\ 
+\Phi ^{4}\rho ^{2}d\rho ^{2}+2\Phi ^{3}\rho ^{3}d\rho d\Phi +\Phi ^{2}\rho
^{4}d\Phi ^{2}].%
\end{array}
\tag{80}
\end{equation}%
It is not difficult to see that this expression can be reduced to

\begin{equation}
B=\frac{1}{1-\Phi ^{2}\rho ^{2}}[2\Phi \rho d\rho d\Phi +\rho ^{2}d\Phi
^{2}++\Phi ^{4}\rho ^{2}d\rho ^{2}].  \tag{81}
\end{equation}%
Using (64), (72) and (73) one discovers that the term in brackets in (81)
leads to

\begin{equation}
\begin{array}{c}
2\Phi \rho d\rho d\Phi +\rho ^{2}d\Phi ^{2}+\Phi ^{4}\rho ^{2}d\rho ^{2}= \\ 
\\ 
=-\frac{16\rho ^{2}d\rho ^{2}}{(\rho ^{2}+1)^{3}}+\frac{16\rho ^{4}d\rho ^{2}%
}{(\rho ^{2}+1)^{4}}+\frac{16\rho ^{2}d\rho ^{2}}{(\rho ^{2}+1)^{4}}= \\ 
\\ 
\frac{16\rho ^{2}d\rho ^{2}}{(\rho ^{2}+1)^{4}}[-(\rho ^{2}+1)+\rho
^{2}+1]=0.%
\end{array}
\tag{82}
\end{equation}%
This result implies that $B=0$. Thus, by means of two different, but
equivalent, methods, we have shown that stereographic transformation (61)
leads to a metric of the form of (68).

It is interesting to calculate the Riemann tensor associated to the metric
(68), namely

\begin{equation}
h_{ij}=\Phi ^{2}\delta _{ij}.  \tag{83}
\end{equation}%
First, by convenience one writes $\Phi =e^{\lambda }$. The corresponding
Christoffel symbols are

\begin{equation}
\Gamma _{kl}^{i}=\delta _{k}^{i}\lambda ,_{l}+\delta _{l}^{i}\lambda
,_{k}-\delta _{kl}\delta ^{il}\lambda ,_{l},  \tag{84}
\end{equation}%
and the Riemann tensor becomes%
\begin{equation}
\begin{array}{c}
R_{jkl}^{i}=\delta _{l}^{i}\lambda ,_{jk}-\delta _{k}^{i}\lambda
,_{jl}-\delta _{jl}\lambda ,_{km}\delta ^{im}+\delta _{jk}\lambda
,_{lm}\delta ^{im} \\ 
\\ 
+(\delta _{k}^{i}\lambda ,_{l}-\delta _{l}^{i}\lambda ,_{k})\lambda
,_{j}-(\delta _{k}^{i}\delta _{jl}-\delta _{l}^{i}\delta _{jk})\lambda
,_{k}\delta ^{kl}\lambda ,_{l} \\ 
\\ 
-(\delta _{jk}\lambda ,_{l}-\delta _{jl}\lambda ,_{k})\delta ^{il}\lambda
,_{l}.%
\end{array}
\tag{85}
\end{equation}%
Therefore the Ricci tensor is%
\begin{equation}
\begin{array}{c}
R_{jl}=-(d-2)\lambda ,_{jl}-\delta _{jl}\lambda ,_{km}\delta
^{km}+(d-2)(\lambda ,_{j}\lambda ,_{l}-\delta _{jl}\lambda ,_{m}\delta
^{mn}\lambda ,_{n} \\ 
\\ 
-(\delta _{jk}\lambda ,_{l}-\delta _{jl}\lambda ,_{k})\delta ^{il}\lambda
,_{l},%
\end{array}
\tag{86}
\end{equation}%
and the scalar curvature gives

\begin{equation}
R=e^{-2\lambda }(-2(d-1)\lambda ,_{km}\delta ^{km}-(d-2)(d-1)\delta
^{mn}\lambda ,_{m}\lambda ,_{n}.  \tag{87}
\end{equation}%
Thus, the Lagrangian

\begin{equation}
L=\frac{1}{2}\sqrt{h}(R+\Lambda ),  \tag{88}
\end{equation}%
leads to

\begin{equation}
\begin{array}{c}
L=-\frac{(d-1)}{(d-2)}\partial _{m}(e^{(d-2)\lambda }),_{n}\delta ^{mn} \\ 
\\ 
+\frac{1}{2}(d-2)(d-1)e^{(d-2)\lambda }\delta ^{mn}\lambda ,_{m}\lambda
,_{n}+\frac{\Lambda }{2}e^{d\lambda },%
\end{array}
\tag{89}
\end{equation}%
where we used the fact that $h=\det h_{ij}=e^{d\lambda }$ and we assume, of
course, that $d-2\neq 0$ and $d-1\neq 0$. The first term in (89) can be
dropped since it is a total derivative. So, (89) is reduced to

\begin{equation}
L=\frac{1}{2}(d-2)(d-1)e^{(d-2)\lambda }\delta ^{mn}\lambda ,_{m}\lambda
,_{n}+\frac{\Lambda }{2}e^{d\lambda }.  \tag{90}
\end{equation}%
Rewriting this expression in terms of the original variable $\Phi
=e^{\lambda }$ one gets

\begin{equation}
L=\frac{1}{2}(d-2)(d-1)\Phi ^{(d-4)}\delta ^{mn}\Phi ,_{m}\Phi ,_{n}+\frac{%
\Lambda }{2}\Phi ^{d}.  \tag{91}
\end{equation}%
Thus, the Euler-Lagrange equations

\begin{equation}
\partial _{i}(\frac{\partial L}{\partial \Phi ,_{i}})-\frac{\partial L}{%
\partial \Phi }=0,  \tag{92}
\end{equation}%
yield%
\begin{equation}
\begin{array}{c}
\partial _{i}[(d-2)(d-1)\Phi ^{(d-4)}\delta ^{in}\Phi ,_{m}] \\ 
\\ 
-\frac{1}{2}(d-4)(d-2)(d-1)\Phi ^{(d-5)}\delta ^{mn}\Phi ,_{m}\Phi ,_{n} \\ 
\\ 
+\frac{\Lambda d}{2}\Phi ^{d-1}=0.%
\end{array}
\tag{93}
\end{equation}%
This expression can be simplified in the form%
\begin{equation}
\begin{array}{c}
\lbrack (d-2)(d-1)\Phi ^{(d-4)}\delta ^{in}\Phi ,_{im}] \\ 
\\ 
+\frac{1}{2}(d-4)(d-2)(d-1)\Phi ^{(d-5)}\delta ^{mn}\Phi ,_{m}\Phi ,_{n} \\ 
\\ 
+\frac{\Lambda d}{2}\Phi ^{d-1}=0.%
\end{array}
\tag{94}
\end{equation}%
Multiplying by $\Phi ^{-(d-4)}$ it can be further simplified to%
\begin{equation}
\begin{array}{c}
\lbrack (d-2)(d-1)\delta ^{in}\Phi ,_{im}]+\frac{1}{2}(d-4)(d-2)(d-1)\Phi
^{-1}\delta ^{mn}\Phi ,_{m}\Phi ,_{n} \\ 
\\ 
+\frac{\Lambda d}{2}\Phi ^{3}=0.%
\end{array}
\tag{95}
\end{equation}

Let us show that (64) is a solution of this equation. Substituting (64) into
(95) one gets%
\begin{equation}
\begin{array}{c}
(d-2)(d-1)\delta ^{in}(-\frac{4\delta _{in}}{(1+X^{k}X_{k})^{2}}+\frac{%
16X_{i}X_{n}}{(1+X^{k}X_{k})^{3}}) \\ 
\\ 
+\frac{4}{(1+X^{k}X_{k})^{3}}(d-4)(d-2)(d-1)(1+X^{k}X_{k})\delta
^{mn}X_{m}X_{n} \\ 
\\ 
+\frac{4\Lambda d}{(1+X^{k}X_{k})^{3}}=0.%
\end{array}
\tag{96}
\end{equation}%
Rewriting this expression

\begin{equation}
\begin{array}{c}
\frac{1}{(1+X^{k}X_{k})^{3}}[(d-2)(d-1)(-4d(1+X^{k}X_{k})+16X^{k}X_{k}) \\ 
\\ 
+4(d-4)(d-2)(d-1)X^{k}X_{k}+4\Lambda d]=0,%
\end{array}
\tag{97}
\end{equation}%
one observes that (97) is consistent if 
\begin{equation}
\Lambda =(d-2)(d-1).  \tag{98}
\end{equation}%
Therefore we have shown that the stereographic projection is consistent with
a theory of non-vanishing cosmological constant $\Lambda $ given in (98).

\bigskip \ 

\noindent \textbf{6. Final remarks}

\smallskip \ 

It is known that some of the most interesting generalizations of spheres are
the squashed spheres in supergravity and pseudo spheres in oriented matroid
theory. The idea of squashed spheres arises in attempt to find a realistic
Kaluza- Klein theory. In fact the typical spontaneous compactification in
eleven dimensional supergravity is given by $M^{4}\times S^{7}$. But $S^{7}$
is isomorphic to $SO(8)/SO(7)$ which implies that instead of the group $%
U(1)\times SU(2)\times SU(3)$ of the standard model, the transition group is 
$SO(8)$. Thus, one uses the concept of spontaneous symmetry braking in order
to make the transition $SO(8)\rightarrow U(1)\times SU(2)\times SU(3)$. One
possibility to achieve this goal is to assume that the symmetry braking
induces the transition $S^{7}\longrightarrow \mathcal{S}^{7}$, where $%
\mathcal{S}^{7}$ is the squashed seven sphere [22]. The simplest example of
this process is provided by the squashed $S^{3}$ sphere [31]. In this case
the original metric

\begin{equation}
ds^{2}=\sigma _{1}+\sigma _{2}+\sigma _{3},  \tag{99}
\end{equation}%
with invariant group $SO(4)$ is broken to the form

\begin{equation}
ds^{2}=\sigma _{1}+\sigma _{2}+\lambda \sigma _{3},  \tag{100}
\end{equation}%
where $\lambda \neq 1$ and $\sigma _{1},\sigma _{2}$ and $\sigma _{3}$ have
quadratic form. When $\lambda =1$ (100) is reduced to the line element of $%
S^{3}$, that is to (99). The idea is to get the reduction $SO(4)\rightarrow
U(1)\times SU(2)$ by the process of symmetry braking.

There are important topological aspects related with the present approach of
higher dimensional spheres. Mathematically, it may be interesting to link
our work with the Bott periodicity theorem (see Ref. [10] and references
therein). Moreover, we would like also to describe an application of
Division-algebra/Poincar\'{e}-conjecture correspondence in qubits theory. It
has been mentioned in Ref. [16], and proved in Refs. [17] and [18], that for
normalized qubits the complex $1$-qubit, $2$-qubit and $3$-qubit are deeply
related to division algebras via the Hopf maps, $S^{3}\overset{S^{1}}{%
\longrightarrow }S^{2}$, $S^{7}\overset{S^{3}}{\longrightarrow }S^{4}$ and $%
S^{15}\overset{S^{7}}{\longrightarrow }S^{8}$, respectively. It seems that
there is not a Hopf map for higher $N$-qubit states. Therefore, from the
perspective of Hopf maps, and therefore of division algebras, one arrives to
the conclusion that $1$-qubit, $2$-qubit and $3$-qubit are more special than
higher dimensional qubits (see Refs. [16]-[17] for details). Considering the 
$2$-qubit as a guide one notices that $S^{3}$ plays the role of fiber in the
map $S^{7}\overset{S^{3}}{\longrightarrow }S^{4}$. Thus, in principle one
may think in a more general map $\mathcal{M}^{7}\overset{\mathcal{M}^{3}}{%
\longrightarrow }\mathcal{M}^{4}$ leading to a more general $2$-qubit system
which one may call $2$-Poinqubit (just to remember that this is a concept
inspired by Poincar\'{e} conjecture.) At the end one may be able to obtain
the transition $2$-Poinqubit$\longrightarrow 2$-qubit. Of course one may
extend most of the arguments developed in this work to the other Hopf maps $%
S^{3}\overset{S^{1}}{\longrightarrow }S^{2}$ and $S^{15}\overset{S^{7}}{%
\longrightarrow }S^{8}$.

Let us now discuss some physical scenarios where the division-algebra/Poincar%
\'{e}-conjecture correspondence may be relevant. Let us start by recalling
the Einstein field equations with cosmological constant $\Lambda $,

\begin{equation}
R_{ij}-\frac{1}{2}\gamma _{ij}R+\Lambda \gamma _{ij}=0.  \tag{101}
\end{equation}%
It is known that the lowest energy solution of (101) corresponds precisely
to $S^{3}$ (or to $S^{d}$ in general). In this case the cosmological
constant $\Lambda $ is given by $\Lambda =\frac{2}{r_{0}^{2}}$. From quantum
mechanics perspective One may visualize $\mathcal{M}^{3}$ as an excited
state which must decay (homeomorphically) to $S^{3}$, according to the
Poincar\'{e} conjecture. Symbolically, one may write this as $\mathcal{M}%
^{3}\rightarrow S^{3}$.

In Ref. [15] it is observed that the transition $\mathcal{M}^{3}\rightarrow
S^{3}$ may be applied in two important scenarios: special relativity and
cosmology. In the first case the evolution process $\varphi
(v_{x},v_{y},v_{z})\rightarrow \sqrt{c^{2}-(v_{x}^{2}+v_{y}^{2}+v_{z}^{2})}$
may be understood as the transition $\mathcal{C}\rightarrow c$ of the light
velocity (see Ref. [15] for details). While in the second case the standard
Friedmann-Robertson-Walker universe corresponds to a time evolving radius of
a $S^{3}$ space. can be modified in $\mathcal{M}^{3}$. Thus, at the end the
acceleration may produce a phase transition changing $\mathcal{M}^{3}$ to a
space of constant curvature which corresponds precisely to the de Sitter
phase associated with $S^{3}$.

Moreover, in Ref. [15] it was proposed the complex generalization

\begin{equation}
i\frac{\partial \psi _{ij}}{\partial t}=-2\mathcal{R}_{ij},  \tag{102}
\end{equation}%
of (57). Here, the metric $\gamma _{ij}$ and the Ricci tensor $R_{ij}$ are
may be complexified, $\gamma _{ij}\rightarrow \psi _{ij}$ and $%
R_{ij}\rightarrow \mathcal{R}_{ij}$, respectively. The idea is now to
consider the evolving complex metric $\psi _{ij}$.

Finally, one may be interested in a possible connection of the Poincar\'{e}
conjecture with oriented matroid theory [32] (see also Refs. [33]-[38] and
references therein). This is because for any sphere $S^{d}$ one may
associate a polyhedron which under stereographic projection corresponds to a
graph in $R^{d+1}$. It turns out that matroid theory can be understood as a
generalization of graph theory and therefore it may be interesting to see
whether there is any connection between oriented matroid theory and Poincar%
\'{e} conjecture. In fact in oriented matroid theory there exist the concept
of pseudo-spheres which generalizes the ordinary concept of spheres (see
Ref. [32] for details). So one wonders if there exist the analogue of Poincar%
\'{e} conjecture for pseudo-spheres.

\bigskip \ 

\begin{center}
\textbf{Acknowledgments}
\end{center}

This work was partially supported by PROFAPI-UAS-2013.

\end{document}